\documentclass[citeautoscript,floatfix,aps,prb,twocolumn,
      superscriptaddress]{revtex4-2}

\usepackage{amsmath}
\usepackage{amssymb}
\usepackage{graphicx}
\usepackage{epstopdf}
\usepackage{natbib}
\usepackage{array}
\usepackage[utf8]{inputenc}
\usepackage{xspace}
\usepackage{multirow}
\usepackage{changes}
\usepackage{dcolumn}
\usepackage{bm}
\renewcommand{\vec}{\boldsymbol} 
\usepackage{braket}
\usepackage{hyperref}
\hypersetup{
colorlinks=true,final=true,
        linkcolor=blue,
        citecolor=blue,
        filecolor=blue,
        urlcolor=blue,
}
\usepackage{cleveref}
\usepackage{changes}

\newcommand{\cubic}{$Pm\overline{3}m$\xspace}
\newcommand{\smo}{SrMoO$_3$}
\newcommand{\pmo}{PbMoO$_3$}
\newcommand{\sro}{SrRuO$_3$}
\newcommand{\svo}{SrVO$_3$}

\begin{document}
\title{Low-temperature transport in high-conductivity correlated metals: A density-functional plus dynamical mean-field study of cubic perovskites}
\author{Harrison LaBollita}
\thanks{These authors contributed equally to this work.}
\affiliation{Center for Computational Quantum Physics,
             Flatiron Institute,
             162 5th Avenue, New York, New York 10010, USA.}
\author{Jeremy Lee-Hand}
\thanks{These authors contributed equally to this work.}
\affiliation{Department of Physics and Astronomy,
             Stony Brook University,
             Stony Brook, New York, 11794-3800, USA}
\author{Fabian B.~Kugler}
\thanks{These authors contributed equally to this work.}
\affiliation{Center for Computational Quantum Physics,
             Flatiron Institute,
             162 5th Avenue, New York, New York 10010, USA.}
\affiliation{Institute for Theoretical Physics, University of Cologne, 50937 Cologne, Germany}
\author{Lorenzo Van Mu\~{n}oz}
\affiliation{Department of Physics,
             Massachusetts Institute of Technology,
             77 Massachusetts Avenue, Cambridge, Massachusetts 02139, USA.}
\author{Sophie Beck}
\affiliation{Center for Computational Quantum Physics,
             Flatiron Institute,
             162 5th Avenue, New York, New York 10010, USA.}
\author{Alexander Hampel}
\affiliation{Center for Computational Quantum Physics,
             Flatiron Institute,
             162 5th Avenue, New York, New York 10010, USA.}
\author{Jason Kaye}
\affiliation{Center for Computational Quantum Physics, Flatiron Institute, 162 5th Avenue, New York, New York 10010, USA.}
\affiliation{Center for Computational Mathematics, Flatiron Institute, 162 5th Avenue, New York, New York 10010, USA.}
\author{Antoine Georges}
\affiliation{Coll{\`e}ge de France, 11 Place Marcelin Berthelot, 75005 Paris, France}
\affiliation{Center for Computational Quantum Physics,
             Flatiron Institute,
             162 5th Avenue, New York, New York 10010, USA.}
\affiliation{CPHT, CNRS, {\'E}cole Polytechnique, IP Paris, F-91128 Palaiseau, France}
\affiliation{DQMP, Universit{\'e} de Gen{\`e}ve, 24 Quai Ernest Ansermet, CH-1211 Gen{\`e}ve, Suisse}
\author{Cyrus E. Dreyer}
\thanks{Contact author, cyrus.dreyer@stonybrook.edu}
\affiliation{Department of Physics and Astronomy,
             Stony Brook University,
             Stony Brook, New York, 11794-3800, USA}
\affiliation{Center for Computational Quantum Physics,
             Flatiron Institute,
             162 5th Avenue, New York, New York 10010, USA.}
\date{\today}

\begin{abstract}
While methods based on density-functional perturbation theory have dramatically improved our understanding of electron-phonon contributions to transport in materials, methods for accurately capturing electron-electron scattering relevant to low temperatures have seen significantly less development. The case of high-conductivity, moderately correlated materials characterized by low scattering rates is particularly challenging, since exquisite numerical precision of the low-energy electronic structure is required. Recent methodological advancements to density-functional theory combined with dynamical mean-field theory (DFT+DMFT), including adaptive Brillouin-zone integration and numerically precise self-energies, enable a rigorous investigation of electron-electron scattering in such materials. In particular, these tools may be leveraged to perform a robust scattering-rate analysis on both real- and imaginary-frequency axes. Applying this methodology to a subset of ABO$_3$ perovskite oxides -- \svo{}, \smo{}, \pmo{}, and \sro{} -- we demonstrate its ability to obtain quantitative convergence of the local electron-electron contributions to the temperature-dependent direct-current resistivity. This combination of numerical techniques offers fundamental insight into the role of electronic correlations in transport phenomena and provides a predictive tool for identifying materials with potential for technological applications. 
\end{abstract}

\maketitle

\section{\label{sec:intro}Introduction}
The combination of density-functional theory and dynamical mean-field theory (DFT+DMFT) has emerged as a powerful method for computing the electronic structure of correlated quantum materials and understanding their properties~\cite{Georges1996,Kotliar:2006}. A key application of this theoretical and computational framework is to describe electronic transport phenomena subject to electron-electron (el-el) interactions. These include the direct-current (dc) resistivity \cite{Deng2016,Radetinac2016,Zhang2022,Abramovitch2023,Abramovitch2024}, optical conductivity \cite{Haule2010,Tomczak2012,Nekrasov2013,Kang2021,Deng2013,Deng2014,Stricker2014,Ahn2022}, and Seebeck coefficient \cite{Tomczak2012,Deng2013,Mravlje2016}. Beyond characterizing a material's intrinsic electronic phase, these properties are crucial in assessing its potential for future technological applications; for example, there is an increasing need for materials with high electrical conductivity for more energy- and heat-efficient electronic devices \cite{Gall2020,Gall2021,Kim2024}. 

One of the most fundamental transport properties of a material is its temperature-dependent dc resistivity, $\rho(T)$. As temperature $T$ is decreased, semiconductors and insulators are characterized by an exponentially increasing $\rho$, while metals exhibit a power-law decrease, with $\rho$ falling to zero in the case of superconductivity. For metals, the exponent of the $\rho(T)$ scaling in a given $T$ range is often used to determine the nature of the scattering that limits the electrical conductivity \cite{Ziman1960}. The canonical picture involves four regimes: (1) $T$-independent residual resistivity caused by disorder and impurities which dominates as $T\rightarrow 0$; (2) a low-$T$ $\rho(T)\propto T^2$ region attributed to dominant Fermi liquid (FL) el-el scattering \cite{Landau1937}; (3) a region at higher $T$ where electron-phonon (el-ph) scattering dominates, characterized by $\rho(T)\propto T^5$ transitioning to (4) $\rho(T)\propto T$ (``Bloch--Gr\"{u}neisen'' behavior \cite{Ziman1960}). This linear behavior of $\rho(T)$ may continue to very high $T$ or saturate around the point where the Mott--Ioffe--Regel criterion is satisfied~\cite{Gunnarsson2003,Hussey2004}. 

The temperature region where a given scattering mechanism is relevant depends on the sample quality and the strength of el-el versus el-ph interactions. Material-specific properties, such as Fermi surface morphology \cite{Kukkonen1978} and complex energy/wavevector dependences of el-el and/or el-ph scattering rates can alter these power laws. This raises challenges in identifying the dominant scattering mechanisms from the $T$ dependence of $\rho(T)$ as well as in identifying more exotic electronic phases with novel transport phenomena. As summarized by Bari\u{s}i\'{c} \textit{et al.}, ``When exploring the properties of a material, the resistivity is the quantity that is often first measured, but last understood.''~\cite{Barisic2013}. 

 To address the challenges in interpreting $\rho(T)$ in metals, quantitative first-principles-based calculations are essential. Recent advances in density-functional perturbation theory combined with the Boltzmann transport equation have led to significant progress in understanding the el-ph contribution to transport~\cite{Giustino2017}. The development of \textit{ab-initio} approaches that capture el-el contributions remains far less mature, with DFT+DMFT~\cite{Kotliar:2006} emerging as one of the key methods. 

While DMFT is often invoked for strongly correlated systems~\cite{Georges1996}, it has recently been shown to also accurately capture the electronic structure of materials with weak to moderate correlations~\cite{Mandal2022,Cappelli2022,Hampel2021,Lechermann2021}. Such materials include metals with high conductivities that are interesting for technological applications. The low magnitude of the scattering rates at low $T$ in these systems poses a significant challenge to DFT+DMFT transport calculations. We demonstrate how recent methodological and computational advancements allow us to meet such challenges: Brillouin-zone integrals can be computed to high accuracy using adaptive integration methods~\cite{Kaye2023}, numerically accurate self-energies are confirmed with ``handshake'' agreement between different DMFT impurity solvers, and relevant transport quantities can be robustly extracted from the electronic self-energy using suitable analysis tools on either the real- or imaginary-frequency axes. 

We deploy our computational framework on a representative subset of ABO$_{3}$ perovskite oxides: \svo{}, \smo{}, \pmo{}, and \sro{}. These materials were chosen because they have a similar atomic and electronic structure but distinct transport properties observed experimentally. \svo{} and \smo{} are moderately correlated high-conductivity metals, with $d^1$ and $d^2$ occupancies of their $t_{2g}$ orbitals, respectively. In fact, \smo{} has the lowest reported room-temperature (RT) resistivity of any perovskite oxide \cite{Nagai2005}. As we demonstrate in our companion paper \cite{CompanionPaper}, resistivity measurements on both \svo{} and \smo{} are consistent with FL $\rho(T)\propto T^{2}$ at low $T$, as well as approximately $T^{2}$ el-ph contributions around RT~\cite{Abramovitch2024,coulter:prep}. \pmo{} has the same formal valence as SrMoO$_{3}$ but a measured $\rho$ three orders of magnitude higher and an unusual sublinear $T$ dependence~\cite{Takatsu2017}. Finally, SrRuO$_{3}$ ($d^{4}$) is strongly correlated and will be used to contrast with the previous materials. It exhibits non-FL transport down to $\sim 10$--$30$~K \cite{Bouchard1972,Wu1993,Antognazza1993,Allen1996,Klein1996,mackenzie1998,Cao2008,Ou2019,Wang2020,Zhao2021,Schreiber2023}. Experimentally, the crystal structure of these oxides is either cubic at all $T$ (\svo{}) or exhibits a cubic to orthorhombic transition as $T$ is lowered. To make direct comparisons between the different materials, we employ the cubic (\cubic{}) crystal structure for all of them.

The focus of this work is on the local el-el contributions to $\rho(T)$. For moderately correlated metals, one expects that el-ph scattering, which is not considered here, dominates at all $T$ relevant for practical applications (i.e., around RT). However, achieving a quantitative understanding of el-el scattering remains crucial---not only for fundamental reasons but also, as demonstrated in Ref.~\citenum{CompanionPaper}, to clarify transport behavior by ruling out el-el scattering as the origin of the observed $\rho$ versus $T$ scaling at RT.

As mentioned above, this is a companion paper to Ref.~\citenum{CompanionPaper}. In Ref.~\citenum{CompanionPaper}, we focus on \svo{} and \smo{} and make a comprehensive comparison between experimental and DFT+DMFT-calculated dc resistivity at low $T$, elucidating the different regimes of $\rho(T)\propto T^2$ in those materials. The focus of this paper is to highlight and describe the methodological advances that render such calculations possible. While making direct comparison to experiments is not our focus, we provide some experimental context for \pmo{} and \sro{} in Sec.~\ref{sec:resistivity}. 

The remainder of this paper is organized as follows. In Sec.~\ref{sec:formalism}, we present the Kubo transport formalism and subsequently derive an expression for $\rho$ using the FL form of the electronic self-energy. In Sec.~\ref{sec:methods}, we describe various components of our computational methodology to compute $\rho$ within DFT+DMFT. In Sec.~\ref{sec:resistivity}, we present our computed $\rho(T)$ for each material. Section~\ref{sec:conclusions} contains a summary of the paper and discusses the implications and outlook.

\section{Transport formalism \label{sec:formalism}}

As is common in DFT+DMFT calculations of the dc resistivity,
we start from the Kubo formula for the conductivity.
In this work, we focus on cubic oxides involving the three degenerate $t_{2g}$ $d$ orbitals. Then, within the DMFT approximation of a local self-energy, 
the self-energy $\Sigma_\omega$ is a scalar (proportional to the unit matrix). It follows that, consistent with the momentum independence of the self-energy and (irreducible) vertex, vertex corrections are absent in DMFT
\cite{Khurana1990,Uhrig1995},
so
\begin{align}
\sigma^{\alpha\alpha'} & =
\pi \int_\omega (-f'_\omega) \int_{\vec{k}} \mathrm{Tr}\, v_{\vec{k}}^\alpha A_{\vec{k}\omega} v_{\vec{k}}^{\alpha'} A_{\vec{k}\omega}
.
\label{eq:Kubo}
\end{align}
Here and throughout the paper, we use atomic units where $k_{\text{B}}=\hbar=e=1$.
In Eq.~\eqref{eq:Kubo}, $\alpha$ and $\alpha'$ are Cartesian directions,  $f'_\omega$ is the derivative of the Fermi function with respect to frequency, and the integrals represent
$\int_\omega = \int_{-\infty}^\infty \mathrm{d} \omega$ and $\int_{\vec{k}} = \int_{\mathrm{BZ}} \frac{\mathrm{d}^d k}{(2\pi)^d}$ with BZ the Brillouin zone and $d$ the dimensionality.
Furthermore, $v_{\vec{k}}^\alpha$ is the DFT band velocity at $k$-point $\vec{k}$ in direction $\alpha$, and $A_{\vec{k}\omega}$ is the spectral function. The latter two quantities are matrices in band and spin space, and the trace runs over bands and spins.

We assume paramagnetism and evaluate the trace in the band basis. In the present cubic $t_{2g}$ scenario, $A_{\vec{k}\omega}$ is diagonal in this basis, and
\begin{align}
\sigma^{\alpha\alpha'} & =
\pi \int_\omega (-f'_\omega) \int_{\vec{k}} \sum_{\sigma \nu \nu'} [v_{\vec{k}}^\alpha]_{\nu\nu'} [A_{\vec{k}\omega}]_{\nu'\nu'} [v_{\vec{k}}^{\alpha'}]_{\nu'\nu} [A_{\vec{k}\omega}]_{\nu\nu}
.
\end{align}
In the coherent regime, where the spectral function is sharp compared to temperature, intraband contributions dominate the dc conductivity, so
\begin{align}
\sigma^{\alpha\alpha} & \approx
\pi \int_\omega (-f'_\omega) \int_{\vec{k}} \sum_{\sigma \nu} [v_{\vec{k}}^\alpha]_{\nu\nu}^2 [A_{\vec{k}\omega}]_{\nu\nu}^2
.
\end{align}

Using $[\Sigma_\omega]_{\nu\nu'}= \delta_{\nu\nu'} \Sigma_\omega$, we can express the spectral function as
\begin{align}
[A_{\vec{k}\omega}]_{\nu\nu}
=
-\frac{1}{\pi} \mathrm{Im} \frac{1}{\omega+\mu-\epsilon_{\vec{k}\nu}-\Sigma_\omega}
,
\end{align}
where $\mu$ is the chemical potential and $\epsilon_{\vec{k}\nu}$ are the DFT eigenvalues. Since $[A_{\vec{k}\omega}]_{\nu\nu}$ depends on $\vec{k}$ only via $\epsilon_{\vec{k}\nu}$, we can write the conductivity ($\sigma = \sigma^{\alpha\alpha}$, $\alpha \in \{x, y, z\}$) as 
\begin{align}
\sigma & \approx
\pi \int_\omega (-f'_\omega) \int_{\epsilon} \Phi(\epsilon) \bigg[ \frac{-1}{\pi} \mathrm{Im} \frac{1}{\omega+\mu-\epsilon-\Sigma_\omega} \bigg]^2
.
\end{align}
Here, $\Phi(\epsilon)$ is the transport function,
\begin{equation}
\Phi(\epsilon)
=
\sum_{\sigma\nu} \int_{\vec{k}} 
[v_{\vec{k}}^\alpha]_{\nu\nu}^2
\delta(\epsilon - \epsilon_{\vec{k}\nu})
,
\label{eq:Phi}
\end{equation}
which encapsulates the band-structure contributions to transport. We can interpret Eq.~\eqref{eq:Phi} as representative of the noninteracting Drude weight (the density of ``free'' electrons available for conduction) for a material when $\epsilon$ is set to the Fermi energy $\epsilon_{\text{F}}$ \cite{Resta2018}. Thus, higher values of $\Phi_{\epsilon_{\text{F}}}$ indicate a higher propensity for conductivity before correlations and scattering processes are considered. 

From this expression, an analytic expression for the low-$T$ FL behavior of the resistivity can be derived. Following Refs.~\citenum{Berthod2013,Georges2021},
we define
$\Gamma_\omega\equiv-2\mathrm{Im}\Sigma_\omega$
and
$E^{\mathrm{F}}_\omega \equiv \mu - \mathrm{Re}\Sigma_\omega$.
Then,
\begin{align}
-\frac{1}{\pi} \mathrm{Im} \frac{1}{\omega+\mu-\epsilon-\Sigma_\omega}
& =
-\frac{1}{\pi} \mathrm{Im} \frac{1}{\omega+E^{\mathrm{F}}_\omega-\epsilon+i\Gamma_\omega/2}
\nonumber \\
& =
\frac{\Gamma_\omega / (2\pi)}{(\omega+E^{\mathrm{F}}_\omega-\epsilon)^2+\Gamma_\omega^2/4}
.
\end{align}
Turning to the energy integral at a fixed value of $\omega$,
we substitute $\epsilon = \omega + E^{\mathrm{F}}_\omega + \frac{1}{2} \Gamma_\omega y$
to arrive at \cite{Georges2021}
\begin{align}
& 
\pi \int \mathrm{d} \epsilon \, \Phi_\epsilon
\bigg[
\frac{\Gamma_\omega / (2\pi)}{(\omega+E^{\mathrm{F}}_\omega-\epsilon)^2+\Gamma_\omega^2/4}
\bigg]^2
\nonumber \\
& =
\frac{2\pi}{\Gamma_\omega}
\int \mathrm{d} y \, \Phi_{\omega+E^{\mathrm{F}}_\omega+\frac{1}{2}\Gamma_\omega y}
\bigg[ \frac{1/\pi}{1+y^2} \bigg]^2
\nonumber \\
& =
\frac{1}{\Gamma_\omega}
\bigg[ \Phi_{\omega+E^{\mathrm{F}}_\omega} + \mathit{O}(\Phi''_{\omega+E^{\mathrm{F}}_\omega}) \bigg]
.
\end{align}
$\Phi_{\omega+E^{\mathrm{F}}_\omega+\frac{1}{2}\Gamma_\omega y}$ in $y$,
using that odd orders vanish under the integral.

In the frequency integral, we expand $\omega+E^{\mathrm{F}}_\omega$ about $\omega=0$, with $\epsilon_{\mathrm{F}} = E^{\mathrm{F}}_0$.
If we further assume $\Gamma_\omega = \Gamma_{-\omega}$, as in the FL regime, odd orders again vanish and we get
\begin{align}
& \int_\omega (-f'_\omega)
\frac{1}{\Gamma_\omega}
\bigg[ \Phi_{\omega+E^{\mathrm{F}}_\omega} + \mathit{O}(\Phi''_{\omega+E^{\mathrm{F}}_\omega}) \bigg]
\nonumber \\
& =
\int_\omega (-f'_\omega)
\frac{1}{\Gamma_\omega}
\bigg[ \Phi_{\epsilon_{\mathrm{F}}} + \mathit{O}(\Phi''_{\epsilon_{\mathrm{F}}}) \bigg]
.
\end{align}
Note that higher orders in the expansion of $\Phi$ lead to higher orders beyond $T^2$ in the expansion of $\rho$ and are thus irrelevant for the present derivation.
The remaining calculation involves
only the scattering rate, which we take in FL form as 
\begin{equation}
2 |\mathrm{Im}\Sigma_\omega|
= \Gamma_\omega = 2C(\omega^2 + \pi^2 T^2)
.
\label{eq:IMSig_FL}
\end{equation}
Upon substituting $x = \omega / T$, the frequency integral yields
\begin{align}
\int_\omega (-f'_\omega)
\frac{1}{\Gamma_\omega}
& =
\frac{1}{8 C T}
\int \mathrm{d} \omega \frac{\cosh^{-2}[\omega/(2T)]}{\omega^2+\pi^2 T^2}
\nonumber \\
& =
\frac{1}{8 C T^2} \int \mathrm{d} x \frac{\cosh^{-2}(x/2)}{x^2+\pi^2}
=
\frac{1}{24 C T^2}
.
\end{align}
Our final result for the dominant contribution to the FL dc resistivity $\rho = 1/\sigma$ is
\begin{align}
\rho & \approx
A T^2
,  \quad
A = \frac{24 C}{\Phi(\epsilon_{\mathrm{F}})}
.
\label{eq:rho_FL}
\end{align}

Thus, we have two options for calculating the FL $\rho(T)$ at low $T$, the full Kubo formula in Eq.~(\ref{eq:Kubo}) or the simplified version of Eq.~(\ref{eq:rho_FL}). While more general, Eq.~(\ref{eq:Kubo}) involves the momentum-resolved real-frequency spectral function at low frequencies, which can be challenging to obtain accurately for materials with low scattering rates. This issue is partially addressed by using adaptive momentum-integration techniques (Sec.~\ref{sec:BZ}), but also stems from problems with analytic continuation of the self-energy at low $\omega$ (Sec.~\ref{sec:trans_FL}). Alternatively, we will show in Sec.~\ref{sec:trans_FL} that the constant $C$ in Eq.~(\ref{eq:rho_FL}) can be obtained from higher-frequency ranges of the FL self-energy utilizing the known FL scaling [Eq.~(\ref{eq:IMSig_FL})]. This makes Eq.~(\ref{eq:rho_FL}) a robust choice for obtaining $\rho$ in high-conductivity materials at low $T$. We compare both options in Sec.~\ref{sec:resistivity} (see Fig.~\ref{fig:rho}).

\begin{figure*}
    \centering
    \includegraphics[width=2\columnwidth]{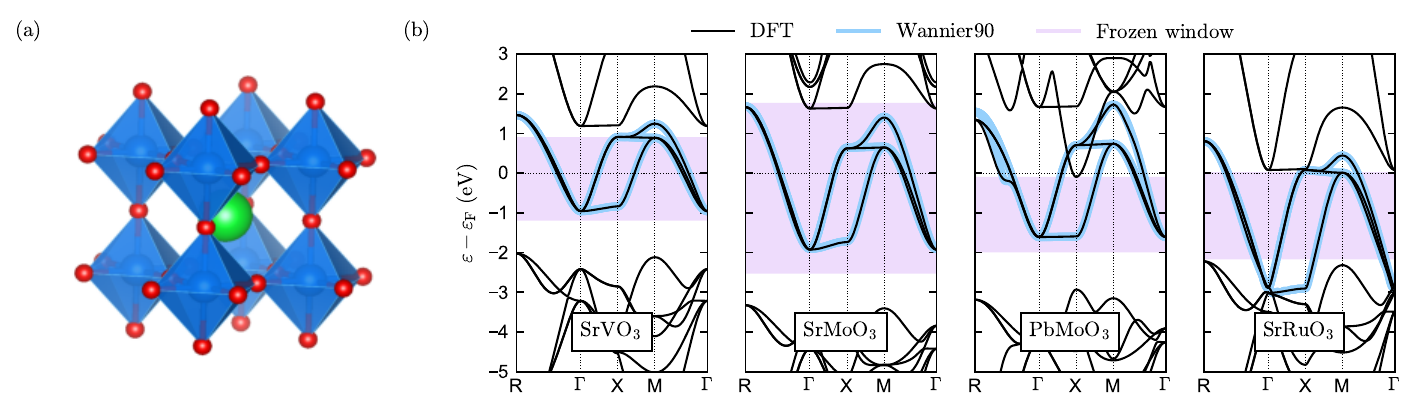}
    \caption{(a) Crystal structure for cubic (\cubic{}) ABO$_{3}$ perovskites, where A, B, and O are denoted as green, blue, and red spheres, respectively. (b) DFT band structures (black lines) compared with Wannier dispersion (light blue) along high-symmetry lines in the Brillouin zone for cubic (\cubic{}) \svo{}, \smo{}, \pmo{}, and \sro{}. Shaded region (light purple) denotes the frozen window used for the downfolding scheme as implemented in Wannier90.}
    \label{fig:bands}
\end{figure*}

\section{\label{sec:methods}Computational Methodology}

In this section, we describe in detail the computational methodologies and analyses required to accurately compute $\rho(T)$ within our DFT+DMFT framework.

\subsection{Interpolating DFT}

For each material, we obtain the optimized crystal structure and converged electron density from DFT as implemented in the Vienna \text{ab-initio} simulation package (VASP)~\cite{Kresse:1993bz,Kresse:1996kl,Kresse:1999dk}. For a complete summary of all computational parameters used, see App.~\ref{app:dft}. Figure~\ref{fig:bands} summarizes the obtained DFT data for: \svo{} ($d^{1}$), \smo{} ($d^{2}$), \pmo{} ($d^{2}$), and \sro{} ($d^{4}$). The bands near the Fermi level are comprised of transition metal (B-site) $t_{2g}$ states with notably different bandwidths, band filling, and hybridization with other bands.

These transition-metal $t_{2g}$ states will serve as the active correlated subspace for our DMFT calculations. We downfold onto these bands by constructing a local atomic-like orbitals using maximally localized Wannier functions (MLWF) \cite{PhysRevB.80.155134} as implemented in Wannier90~\cite{Mostofi_et_al:2014}.  The band structure obtained from our Wannier functions is compared to the Kohn--Sham DFT band structure in Fig.~\ref{fig:bands}(b). The agreement demonstrates that our Wannier functions are a faithful representation of the DFT bands near the Fermi level. To obtain a consistent description for all four materials, we have disentangled the $t_{2g}$ bands from any other hybridizing DFT bands. Table~\ref{tab:wann_windows} provides the window parameters used for each material. The width of the ``frozen windows,'' where exact energy agreement is enforced, are indicated in light purple in Fig.~\ref{fig:bands}(b). 

\begin{table}[b]
 \renewcommand*{\arraystretch}{1.4} 
 \caption{Energy windows for the calculation of Wannier functions. All energies are given  in eV and relative to the Fermi level for each material.}\label{tab:wann_windows}
\begin{ruledtabular}
    \centering
\begin{tabular}{l|cc|cc}
        & \multicolumn{2}{c|}{Frozen} & \multicolumn{2}{c}{Disentanglement}  \\ \hline
SrVO$_3$  & $0.907$               & $-1.192$                                                    & $3.007$                                                    & $-1.292$                                                   \\
SrMoO$_3$ & $1.768$               & $-2.531$                                                    & --                                                        & --                                                        \\
PbMoO$_3$ & $-0.094$              & $-1.994$                                                    & $3.705$                                                    & $-3.494$                                                   \\
SrRuO$_3$ & $0.028$               & $-2.171$                                                    & $0.828$                                                    & $-3.171$                                                  
\end{tabular}
    \end{ruledtabular}
\end{table}

In addition to providing local atomic orbitals for DMFT calculations, ``Wannierization'' of the DFT bands enables efficient interpolation of the electronic structure onto dense reciprocal space meshes. This allows for accurate calculation of the Brillouin-zone (BZ) integrals appearing in the dc conductivity [Eq.~\eqref{eq:Kubo}]. The most common integration method is simple summation over a uniform grid \cite{Pizzi_et_al:2020,wang2006,Tsirkin2021,Kaye2023}. However, for moderately correlated materials (like \svo{}, \smo{}, \pmo{}), the electron-electron scattering rates at low $T$ are on the order of 1~meV, potentially requiring billions of $\vec{k}$ quadrature points per $\omega$ quadrature point for precise calculations \cite{vanmunoz24}. In the present case, one needs at most $200^3$, i.e., 8 million points for percent-level convergence. Several alternative methods have been proposed~\cite{wang2006,Tsirkin2021,assmann16,vanmunoz24,Haule_et_al:2005,oudovenko06,ambrosch2006,methfessel89,bjorkman11,Yates_et_al:2007,chen22,Kaye2023,duchemin23,letournel24}, see the introductions of Refs.~\onlinecite{Kaye2023,vanmunoz24} for further discussion. To access small scattering rates with controlled accuracy in our calculations of the conductivity, we use the recently developed iterated adaptive integration (IAI) method \cite{Kaye2023}, described in Ref.~\onlinecite{vanmunoz24} and implemented in the \textsc{AutoBZ.jl} package \cite{Van-Munoz/Beck/Kaye:2024}. This method reduces the $O({\eta^{-3}})$ scaling of each BZ integral in the dc conductivity using the uniform grid approach to $O[{\log^3(\eta^{-1})}]$ (where $\eta$ is a broadening parameter), automates convergence to a user-specified numerical precision, and also treats the $\omega$ integral adaptively.

We also use the IAI method to compute the transport function [Eq.~\eqref{eq:Phi}] at the Fermi level, $\Phi(\epsilon_{\text{F}})$. Since IAI is designed for BZ integrals with non-zero broadening $\eta$, this requires an artificially broadened integrand and extrapolation to the limit $\eta=0$.
Here, we use a Lorentzian broadening in the integrand,
\begin{align}
\Phi(\omega) & =
\lim_{\eta \to 0} 2\pi\eta
\int_{\vec{k}}
\mathrm{Tr}\,
v^\alpha_{\vec{k}} A^\eta_{\vec{k}\omega} v^\alpha_{\vec{k}} A^\eta_{\vec{k}\omega}
,
\end{align}
where $A^\eta_{\vec{k}\omega}$ is the broadened noninteracting spectral function, which can be evaluated in the orbital basis as
\begin{align}
A^\eta_{\vec{k}\omega} 
=
-\tfrac{1}{\pi} \mathrm{Im}\, G^\eta_{\vec{k}\omega} 
, \qquad
G^\eta_{\vec{k}\omega} =
[ \omega + \mathrm{i} \eta - h_{\vec{k}} ]^{-1}
.
\end{align}
Agreement with Eq.~\eqref{eq:Phi} can be verified from the relation
\begin{equation}
[A^\eta_{\vec{k}\omega}]_{\nu\nu}
[A^\eta_{\vec{k}\omega}]_{\nu'\nu'}
\approx
\frac{\eta}{\pi}
\frac{\delta(\omega - \epsilon_{\vec{k}\nu}) + \delta(\omega - \epsilon_{\vec{k}\nu'})}{(\epsilon_{\vec{k}\nu} - \epsilon_{\vec{k}\nu'})^2 + (2\eta)^2}
,
\end{equation}
implying $[A^\eta_{\vec{k}\omega}]_{\nu\nu}^2 \approx
\frac{1}{2\pi\eta}
\delta(\omega - \epsilon_{\vec{k}\nu})$.
Interband contributions, absent in Eq.~\eqref{eq:Phi}, are thus suppressed by $\mathit{O}(\eta^2)$, and the desired result follows in the limit $\eta \to 0$.

\begin{figure}
\centering
\includegraphics[width=\linewidth]{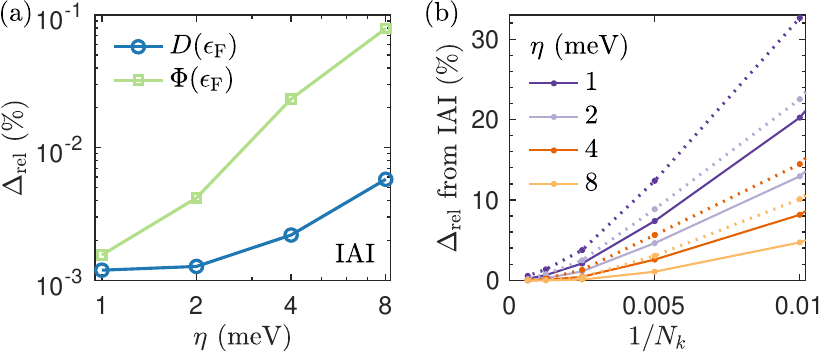}
\caption{Convergence of \smo\ DFT calculated properties at the Fermi level $\epsilon_{\mathrm{F}}$.
(a) DOS and transport function for various broadening $\eta$, obtained with the iterated adaptive integrator (IAI). Relative differences are taken from the values $V_{\mathrm{uc}}D(\epsilon_{\mathrm{F}}) = 1.886$/eV, $\Phi(\epsilon_{\mathrm{F}}) = 3.930\,$eV/($\Omega\,$cm).
(b) DOS (solid lines) and transport function (dotted lines) at each value of $\eta$ computed with a fixed number of $k$ points, $N_k$, per dimension in the full BZ. Differences are taken from the IAI result at the respective value of $\eta$. 
}
\label{fig:wannier_transport_convergence}  
\end{figure}

\subsection{Brillouin-zone integration \label{sec:BZ}}

\begin{figure}
    \centering
    \includegraphics[width=\linewidth]{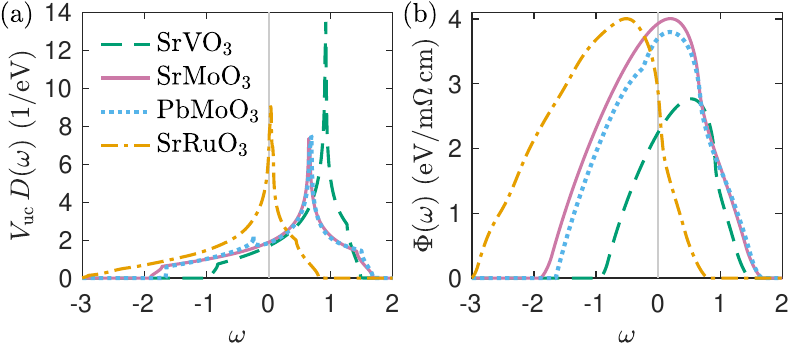}
    \caption{(a) Density of the $t_{2g}$ states and  (b) transport function of all four materials using IAI with $\eta=1$ meV.}\label{fig:wannier_transport}  
\end{figure}

Figure~\ref{fig:wannier_transport_convergence}(a) demonstrates the convergence of the DOS and transport function at the Fermi level with respect to $\eta$, compared with the result at $\eta=1$ meV rounded to three digits (as used in Ref.~\citenum{CompanionPaper}). We observe that convergence to below $10^{-3}$ ($10^{-4}$) is achieved at $\eta=8$ meV ($\eta=2$ meV) for the DOS (transport function). In Fig.~\ref{fig:wannier_transport_convergence}(b), we plot the error of the uniform integration method for the DOS, implemented in \textsc{AutoBZ.jl}, versus the number $N_k$ of quadrature points per dimension in the full BZ. Convergence to below $10^{-2}$ requires about $N_k = 200$ ($N_k = 300$) at $\eta=8$ meV and about $N_k = 700$ ($N_k = 1000$) at $\eta=1$ meV for the DOS (transport function). Note that this requires $N_{k}^{3}$ grid points to resolve the BZ integral. Finally, we plot the calculated $t_{2g}$ DOS and transport function for our oxide materials using IAI with $\eta=1$ meV in Fig.~\ref{fig:wannier_transport}.

\subsection{cRPA interaction parameters: successes and failures}

The next step is to determine the interaction parameters that define the Hubbard--Kanamori Hamiltonian of the correlated $t_{2g}$ impurity problem identified above. Ideally, these parameters are derived from first principles, avoiding any empirical input. The constrained random phase approximation (cRPA) \cite{Aryasetiawan2004} is the standard approach to estimate the static interaction parameters from DFT. For our test case oxides, we compute cRPA  interaction parameters utilizing the implementation in the VASP code~\cite{Kaltak2015} by averaging the full four-index interaction tensor to parametrize the Hubbard--Kanamori Hamiltonian~\cite{vaugier2012}. By the cubic symmetry within the $t_{2g}$ subspace, these parameters reduce to a single Hubbard $U$ and Hund's coupling $J$ parameter, which are summarized for each material in Table~\ref{tab:crpa}.

\begin{table}[b]
  \caption{Hubbard--Kanamori parameters calculated using cRPA for each material (in units of eV). If given, values in parentheses correspond to the empirically chosen ones used in the calculation. We fix $U^{\prime}=U-2J$.}\label{tab:crpa}
 \begin{ruledtabular}
    \centering
\begin{tabular}{ l | l | l }
         & $U$ & $J$ \\ \hline
SrVO$_3$            & 3.32 (4.5) & 0.46 (0.65)       \\
SrMoO$_3$ & 3.07       & 0.31       \\
PbMoO$_3$           & 2.82       & 0.34       \\
SrRuO$_3$           & 2.35       & 0.26 (0.4) \\
\end{tabular}
 \end{ruledtabular}
 \end{table}

\begin{figure*}
    \centering
    \includegraphics[width=2\columnwidth]{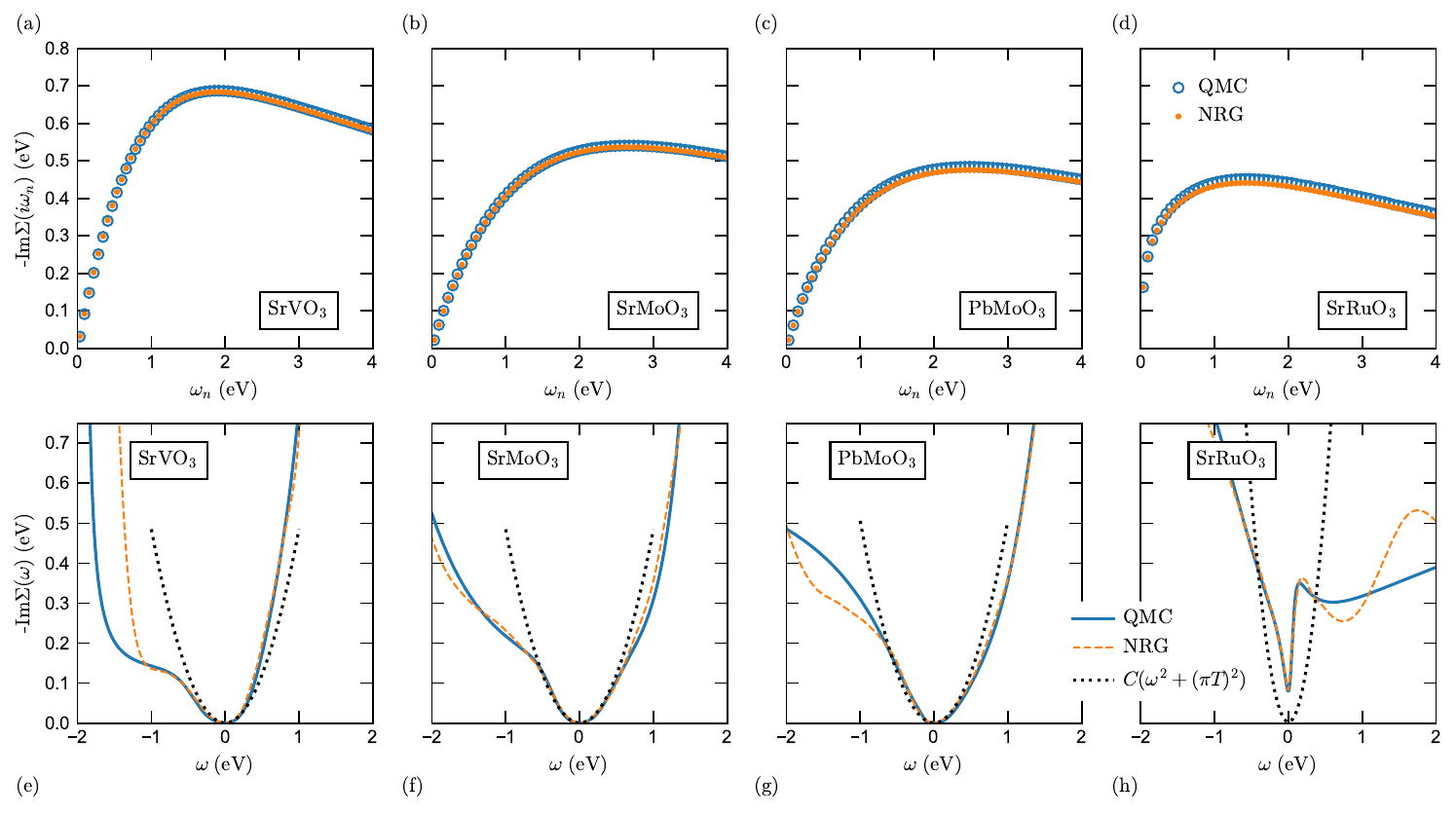}
    \caption{QMC (blue) and NRG (orange) comparison of the DMFT self-energy on the Matsubara imaginary-frequency axis (top) and the real-frequency axis (bottom) for \svo{}, \smo{}, \pmo{}, and \sro{} at $T = 116$ K ($\beta = 100$/eV). The dashed (black) lines indicate a fit of the real-frequency data to the Fermi-liquid form $C(\omega^2 + \pi^2 T^2)$.
    The Fermi-liquid fit to \sro{} is clearly not successful. We note that increasing $C$ to match $\mathrm{Im}\,\Sigma(0)$ does not lead to an overall better fit.}
    \label{fig:sigma-handshake}
\end{figure*}

Interaction parameters obtained from cRPA often provide reliable estimates of the screened interactions in a material. For example, the $U$ and $J$ values computed for \smo{} have been shown to yield results in good agreement with experimental spectroscopy~\cite{hampel2020, Cappelli2022}. Due to the lack of experimental data for \pmo{}, we fully rely on the cRPA values for this compound.


Despite its successes in many cases, cRPA can sometimes fail to capture key experimental observations; this is the case for \svo{} and \sro{}. For \svo{}, experiments report a quasiparticle renormalization factor of $Z\sim0.5$~\cite{Yoshida2005}. In DMFT, this quantity can be estimated from the self-energy as $Z^{-1}=1-\partial_{\omega}\mathrm{Re}\Sigma(\omega,T)|_{\omega\rightarrow0}$. Using the cRPA-derived paramaters in DMFT, we obtain $Z\sim0.6$, while increasing $(U,J)$ to (4.5, 0.65) yields $Z=0.5$, in agreement with experiment~\cite{Lechermann2006}. This artificial increase of the static Hubbard interaction to correctly capture $Z$ has been discussed previously~\cite{Lechermann2006, Taranto2013}. It can be circumvented by including dynamical interactions in combination with non-local correlation effects~\cite{Huang2012,Sakuma2013, Taranto2013}.

Fol \sro{}, the cRPA-predicted interaction parameters fail to capture the correct behavior with respect to the ferromagnetic transition. This transition is especially sensitive to $J$. For the cRPA value of 0.26 eV, we do not observe a ferromagnetic transition down to the lowest temperature studied, while increasing the value to 0.40 eV results in a transition around 290 K. Experimentally, the value is around 150-170 K \cite{Bouchard1972,Eom1992,Wu1993,Antognazza1993,Allen1996,Klein1996,mackenzie1998,Cao2008,Ou2019,Wang2020,Zhao2021,Schreiber2023}, this overestimation is expected since DMFT only includes local magnetic fluctuations and neglects their spatial dependence~\cite{Georges1996}---see, e.g., Ref.~\citenum{Lichtenstein2001} for DMFT estimates of the Curie temperature of transition metals. We note that the interaction parameters are significantly constrained by the fact that they should be roughly the same for \sro{} and CaRuO$_3$, and produce a ferromagnetic (paramagnetic) ground state at low temperatures for \sro{} (CaRuO$_3$) \cite{Dang2015}. This is indeed satisfied for our choice of $U=2.35$ eV and $J=0.4$ eV, see Ref.~\citenum{Dang2015}.

\subsection{Electron-electron self-energies from DMFT} \label{sec:sigma}
We solve the quantum impurity model for the $t_{2g}$ correlated subspace of each material to obtain the local self-energy from el-el interactions within DMFT. The dc resistivity depends sensitively on the low-frequency behavior of $\Sigma(\omega)$ [see Eq.~(\ref{eq:Kubo})] requiring accurate real-frequency data. Commonly used impurity solvers are based on continuous-time quantum Monte Carlo (CT-QMC), which operates exclusively in imaginary time. To obtain real-frequency data for Eq.~(\ref{eq:Kubo}), one must analytically continue from the imaginary axis to the real axis, which is an ill-conditioned mathematical problem \cite{Jarrell1996} and therefore prone to errors and artifacts.

There has been significant progress in analytic continuation (AC) over the past years~\cite{Gunnarsson2010,Fuchs2010,Burnier2013,Sandvik2016,Schott2016,Bao2016,Bergeron2016,Levy2016,Arsenault2017,Otsuki2017,Kraberger2017,Goulko2017,Sim2018,Krivenko2019,Rumetshofer2019,Fei2021_2,Fei2021,Ying2022,Yao2022,Huang2022,Han2022,Sun2023,Huang2023,Huang2023_2,Shao2023,Zhang2024,Zhang2024_2,Khodachenko2024}. However, it is still challenging to systematically ensure accuracy; we show below that low frequencies for high-conductivity materials are particularly problematic. Also, unlike in, e.g., some model systems, a ``ground truth'' solution is almost never available for the self-energies of real materials, precluding systematic benchmarking. 

To address this, we compare the results of two complementary quantum impurity solvers: an imaginary-time CT-QMC-based solver in the hybridization expansion (see App.~\ref{app:QMC}), referred to as QMC, and a real-frequency solver given by the numerical renormalization group (NRG, see App.~\ref{app:NRG}). For the former, we perform AC using the Pad\'e approximate method \cite{FerrisPrabhu1973,Vidberg1977}. The advantage of NRG in this context is clearly that it avoids the above-mentioned uncertainty resulting from AC. However, NRG is limited in the systems it can treat. Increasing the degrees of freedom beyond three spinful orbitals has not been achieved yet, and 
decreasing symmetries (such as breaking spin symmetry) significantly decreases efficiency and/or accuracy. 
Therefore, QMC remains the workhorse for more general impurity problems, and establishing a reliable procedure to extract scattering rates from QMC is essential.

As shown in Fig.~\ref{fig:sigma-handshake}, we obtain a ``handshake'' agreement between the two impurity solvers on both the imaginary- (top row) and real-frequency axes (bottom row) for all materials. \svo{}, \smo{}, and \pmo{} all show the characteristic FL scaling, i.e., $-\mathrm{Im}\Sigma(\omega)\propto\omega^{2} + \pi^2 T^2$ at low energies, while \sro{} clearly exhibits non-FL behavior at this temperature ($T=116$ K). For QMC, this agreement gives us confidence in the convergence of the imaginary-frequency self-energies as well as the AC (at least at relatively low frequencies). For NRG, this system (three orbitals, full Hubbard--Kanamori Hamiltonian with spin-flip and pair-hopping terms) is a considerable challenge, and resolving the FL behavior in $\mathrm{Im}\Sigma$ has only recently become possible \cite{Kugler2022}. Thus, the agreement in Fig.~\ref{fig:sigma-handshake} with the QMC data also gives confidence that the NRG self-energies are accurate.

\subsection{Fermi-liquid regime \label{sec:FL}}

\begin{figure}
    \centering
    \includegraphics[width=\linewidth]{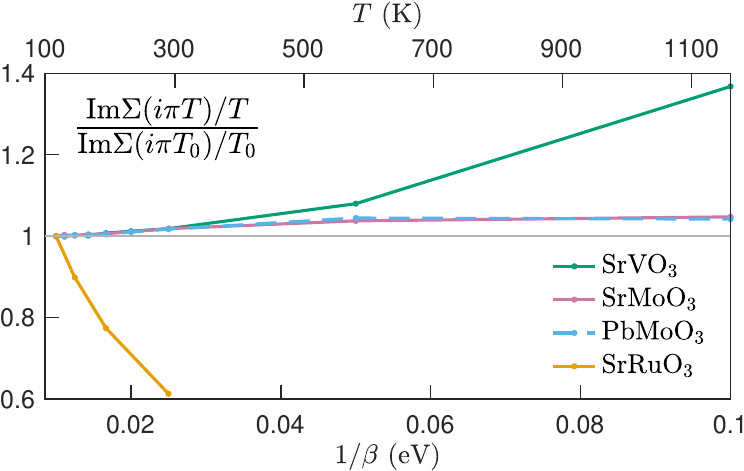}
    \caption{QMC results for the imaginary part of the self-energy at the first Matsubara frequency as a function of $T$. Fermi-liquid behavior indicated by $\mathrm{Im}\Sigma(i\pi T)/T = \mathrm{const}$ \cite{Chubukov2012} is seen below a crossover scale of $T \sim 500\,$K for \svo{}, \smo{}, \pmo{}, but does not occur above $T_0 \approx 116\,$K for \sro{}.
    }\label{fig:Maslov}  
\end{figure}

While fitting Im$\Sigma(\omega)$ to the FL form shown in Fig.~\ref{fig:sigma-handshake} gives a qualitative picture of whether the material is a FL at a given $T$, a more direct approach is to analyze the self-energy on the imaginary-frequency axis. The FL form of the Matsubara self-energy as a function of $i\omega_n \!=\! i(2n \!+\! 1)\pi T$, $n \!\in\! \mathbb{Z}$, is \cite{CompanionPaper}
\begin{equation}
\label{eq:SigMats_FL}
\text{Im}\Sigma(i\omega_n) 
\simeq
(1 - 1/Z) \omega_n + \text{sgn}(\omega_n) C (\omega_n^2-\pi^2 T^2 )
,
\end{equation}
where $Z$ is the quasiparticle weight and $C$ is used to obtain the $A$ coefficient, see Eq.~(\ref{eq:rho_FL}). The FL scale can be quantified by examining $\mathrm{Im}\Sigma(i\omega_0 = i \pi T)$ as a function of $T$ \cite{Chubukov2012}. Indeed, according to Eq.~\eqref{eq:SigMats_FL}, $\mathrm{Im}\Sigma(i \pi T) = (1-1/Z) \pi T + \mathit{O}(T^3)$ in the FL regime, i.e., a linear behavior with $T$. In Fig.~\ref{fig:Maslov}, we plot $\mathrm{Im}\Sigma(i \pi T) / T$ for all four materials. \svo{}, \smo{}, and \pmo{} show rather flat $\mathrm{Im}\Sigma(i \pi T) / T$ (particularly \smo{} and \pmo{}). The deviations in $\mathrm{Im}\Sigma(i \pi T) / T$ from the value at the lowest temperature are on the percent level below a crossover scale of $T_{\mathrm{FL}} \sim 500\,$K. By contrast, \sro{} is nowhere near $\mathrm{Im}\Sigma(i \pi T) / T \sim \mathrm{const}$ in the temperature range considered, implying that $T_{\mathrm{FL}} \ll T_0 \approx 116\,$K.

\begin{figure}
    \centering
    \includegraphics[width=\linewidth]{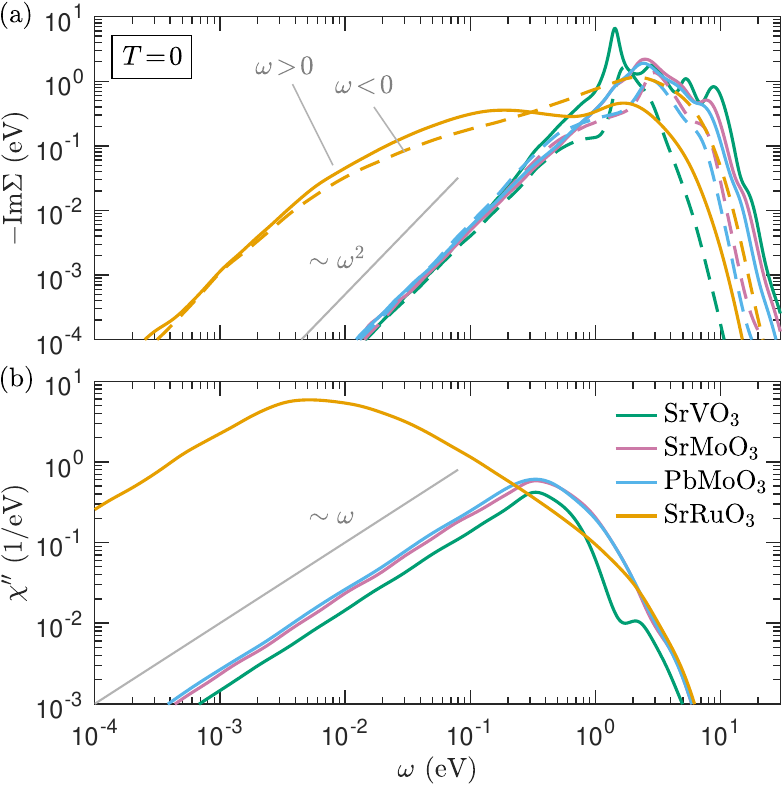}
    \caption{NRG results for the imaginary part of (a) the self-energy and (b) the spin susceptibility at $T = 0$. Fermi-liquid behavior below $\omega \sim 0.1\,$eV for \svo{}, \smo{}, \pmo{} and below $\omega \sim 1\,$meV for \sro{} (in the present paramagnetic and cubic model) is seen by the $\omega^2$ of $\mathrm{Im}\Sigma$ and the $\omega$ scaling of $\chi^{\prime\prime}$. Solid (dashed) lines in (a) are for $\omega>0$ ($\omega<0$).
    }
    \label{fig:Sigma_chi}  
\end{figure}

An obvious benefit of NRG is that results can readily be obtained in real frequencies and at zero temperature. In this case, FL behavior in the self-energy is characterized by $\mathrm{Im}\Sigma \propto \omega^2$. Additionally, it may also be characterized by the dynamic magnetic susceptibility which obeys $\chi^{\prime\prime} \propto \omega$, i.e., linear behavior in the imaginary part. Figures~\ref{fig:Sigma_chi}(a) and (b) show $\mathrm{Im}\Sigma$ and $\chi^{\prime\prime}$, respectively. FL behavior in \svo{}, \smo{}, and \pmo{} is seen below a crossover scale of $\omega_{\mathrm{FL}} \sim 0.1\,$eV. This is consistent with the above estimate if one connects the frequency and temperature coherence scales according to the self-energy's $\omega^2 + \pi^2 T^2$ scaling as $\omega_{\mathrm{FL}} \sim \pi T_{\mathrm{FL}}$. Moreover, our $T=0$ NRG results allow us to infer the value of the FL scale of \sro{} in its (putative) paramagnetic cubic phase (we recall that this material is actually a ferromagnet at low $T$, see Sec.~\ref{sec:resistivity} for further discussion). We find $\omega_{\mathrm{FL}} \sim 1\,$meV ($T_{\mathrm{FL}}\sim 4$ K).

\subsection{Utilizing Fermi-liquid scaling for extracting transport properties \label{sec:trans_FL}}

\begin{table}
\renewcommand*{\arraystretch}{1.4} 
\caption{Extracting the scattering rates in units of 1/eV by polynomial fits of order $p$ to the first $N$ Matsubara frequencies of the QMC result. Here, $C^{(2)}=a_2$ is the coefficient of the quadratic term, and $C^{(0)}=-a_0/(\pi^2 T^2)$ is the coefficient of the constant term divided by $\pi^2 T^2$ where $T\approx 116\,$K. In the Fermi liquid \svo{}, we find $C^{(2)}$ to be stable, while $C^{(0)}$ is unstable due to its small magnitude, giving unphysical negative values. In the non-Fermi liquid \sro{}, $C^{(0)}$ is sizable and thus stable when fitted.
}\label{tab:poly_fits}
\begin{ruledtabular}
\centering
\begin{tabular}{l|c|c|c|c}
$(p,N)$ & $(4,6)$ & $(3,5)$ & $(2,4)$ & $(4,5)$
\\ 
\hline
SrVO$_3$ $C^{(2)}$ & $0.34$ & $0.39$ & $0.43$ & $0.34$
\\
SrVO$_3$ $C^{(0)}$ & $-0.031$ & $-0.10$ & $-0.20$ & $-0.021$
\\
SrRuO$_3$ $C^{(0)}$ & $100$ & $106$ & $117$ & $97$
\\
\end{tabular}
\end{ruledtabular}
\end{table}

The excellent agreement on the scale of Fig.~\ref{fig:sigma-handshake} between the two impurity solvers is not sufficient to guarantee accuracy in $\rho$ for high-conductivity metals. This is because, as mentioned above, we require accuracy of the self-energies on the meV level for the low frequencies needed for transport. We demonstrate here that we can exploit the FL scaling of the self-energy (either Matsubara or real-frequency) to ensure reliable extraction of the scattering rate. Of course, this strategy will not work if the accessible $T$ are outside of the FL regime, as is the case for QMC on \sro{}; in that case, however, we find that the relatively large magnitude of the self-energy at low frequencies alleviates the errors suffered by the higher-conductivity metals. 

The FL coefficients $Z$ and $C$ are often extracted from the Matsubara self-energy by a polynomial fit,
$\text{Im}\Sigma(i\omega_n > 0) 
= \sum_{i=1}^p a_i \omega_n^i$,
of order $p$ to the first $N$ Matsubara frequencies. Comparing this to Eq.~\eqref{eq:SigMats_FL}, we have $a_1 = 1-1/Z$ and two estimates for $C$: $C^{(2)} = a_2$ as well as $C^{(0)}=-a_0/(\pi^2 T^2)$. We find that obtaining $C$ from polynomial fits can be sensitive when the absolute magnitude of $\mathrm{Im}\Sigma(i\omega_{n})$ is small at low frequencies. Table~\ref{tab:poly_fits} summarizes the dispersion of coefficients that can be obtained from polynomial fits by varying the polynomial order $p$ and the number of Matsubara points used in the fit $N$. We use \svo{} and \sro{} as two exemplary cases (\smo{} and \pmo{} are similar to \svo{}). We find that the coefficient $C^{(2)}$ is stable to the choice of $(p,N)$, while $C^{(0)}$ is not if $\mathrm{Im}\Sigma(i\omega_{n})$ is small at low frequencies (the case for \svo{}). Yet, if $\mathrm{Im}\Sigma(i\omega_{n})$ is sufficiently large at low frequencies (the case for \sro{}), $C^{(0)}$ becomes stable to the choice of $(p,N)$, although $C$ alone has limited utility in the non-FL regime. 

In lieu of polynomial fitting in the FL regime, one may also use the following two estimators for the FL coefficients
\cite{CompanionPaper},
\begin{equation}
\label{eq:SigMats_FL_formulas}
1-\frac{1}{Z} = \frac{\text{Im}\Sigma(i\pi T)}{\pi T}
, \ \
C = \frac{\text{Im}\Sigma(3i\pi T)-3\text{Im}\Sigma(i\pi T)}{8\pi^2 T^2}.
\end{equation}
This estimator highlights the difficulty of accurately extracting $C$. At $T = 116$ K, e.g., the numerator is $\sim 10^{-3}$ for \smo{}, \pmo{}, and \svo{}. Consequently, determining $C$ demands accuracy in the first two Matsubara frequencies to at least the fourth decimal place, a level of precision higher than can be discerned in Fig.~\ref{fig:sigma-handshake}.

\begin{figure}
    \centering
    \includegraphics[width=\columnwidth]{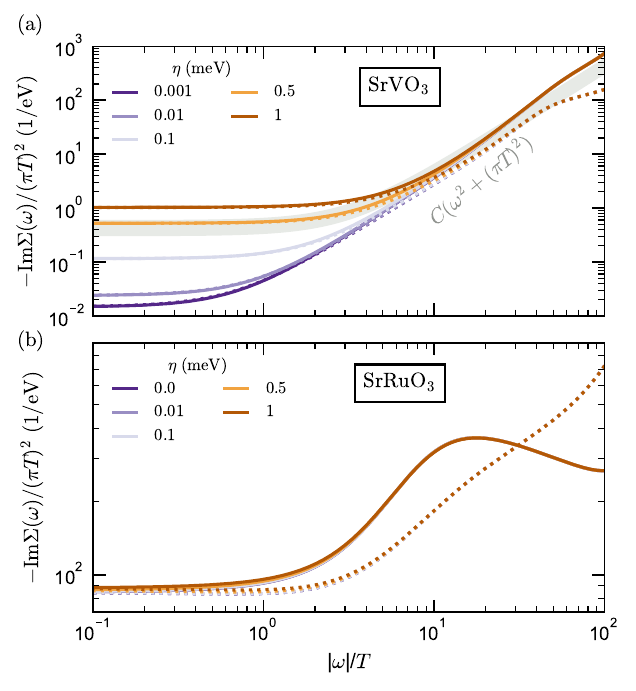}
    \caption{Real-frequency self energies $\Sigma(\omega)$ obtained from Pad\'e analytic continuation for (a) SrVO$_{3}$ and (b) SrRuO$_{3}$ at $T=116$ K plotted as $-\text{Im}\Sigma(\omega)/(\pi T)^{2}$ versus $|\omega|/T$. $\eta$ denotes the frequency offset when evaluating the Pad\'e approximats ($\Sigma(\omega + i\eta)$). The first 400 Matsuara data points were used to construct the Pad\'e approximats. Solid (dashed) lines denote $\mathrm{Im}\Sigma(\omega)$ for $\omega > 0$ ($\omega < 0$). Note $\eta$ is units of meV.}
    \label{fig:pade}
\end{figure}

On the real-frequency axis, the FL form of the self-energy is defined in Eq.~\eqref{eq:IMSig_FL}. In principle, the coefficient $C$ can be taken directly from evaluating $\mathrm{Im}\Sigma(\omega=0)$. However, when using imaginary-time impurity solvers requiring analytic continuation, we will demonstrate that evaluating $\mathrm{Im}\Sigma(\omega=0)$ can be quite sensitive to the continuation. As mentioned above, we choose the Pad\'e analytic continuation to continue the Matsubara data to the real-frequency axis, which is simpler than other popular methods, like the maximum entropy method \cite{Silver1990,Gubernatis1991,Jarrell1996,Bergeron2016,Levy2016,Kraberger2017,Sim2018}, and the community standard for obtaining quality self-energy data at low frequencies \cite{Vidberg1977,Jarrell1996}. This method has a broadening parameter $\eta$, which is used to evaluate the Pad\'e approximants at $\Sigma(\omega + i\eta)$. For moderately correlated materials, the small el-el scattering rate makes the AC results sensitive to this choice. Figure~\ref{fig:pade}(a) demonstrates this for the case of \svo{} (the behavior of \smo{} and \pmo{} is analogous), where we plot $-\text{Im}\Sigma(\omega)/(\pi T)^2$ versus $\vert\omega\vert/T$. Decreasing $\eta$ continuously decreases the magnitude of $\mathrm{Im}\Sigma(\omega=0)$ to clearly unphysical levels. This not only makes extracting $C$ impossible, but is also problematic for using $\mathrm{Im}\Sigma(\omega)$ in the full Kubo formula Eq.~(\ref{eq:Kubo}), which is sensitive to the low-$\omega$ behavior of the self-energy. We can notice in Fig.~\ref{fig:pade}(a) that, for $\vert\omega\vert/T \gtrsim 5$, there is a region where $\mathrm{Im}\Sigma(\omega)$ is insensitive to $\eta$. Since we know the $\omega$ scaling of the FL self-energy [Eq.~\eqref{eq:IMSig_FL}], we can use this region, dominated by the $\omega^2$ behavior, to fit $C$ and extrapolate to small $\omega$. The result of the fitting is given by the gray region in Fig.~\ref{fig:pade}(a), where the finite width denotes our estimated uncertainty in the fit. With this approach, we can obtain a  robust value of $C$, and determine the correct behavior of $\mathrm{Im}\Sigma(\omega)$ at low frequencies.
For $T$ outside of the FL regime, we generally do not know the $\omega$ scaling of $\mathrm{Im}\Sigma(\omega)$ and cannot use this method. However, in such materials with stronger correlations, the magnitude of $\mathrm{Im}\Sigma(\omega=0)$ is usually significantly larger, as is the case for \sro{} [Fig.~\ref{fig:sigma-handshake}(d)]. We show in Fig.~\ref{fig:pade}(b) that this results in a straightforward convergence of the low-$\omega$ behavior of the self-energy with $\eta$, so the AC in this region is expected to be reliable. 

A few additional comments are in order about the extraction of $C$. First, we note that the issues with analyzing the Matsubara self-energy mirror those of the real-frequency self-energy. For imaginary frequencies, the problematic $C^{(0)}$ estimate for $C$ is analogous to \textit{extrapolating} the Matsubara data to $i\omega=0^+$. In a sense, taking the analytically continued data at low frequency is like performing this same extrapolation, since there are no Matsubara data below $\pi T$. By contrast, the $C^{(2)}$ estimate utilizes the higher-frequency data to obtain $C$, which is more robust for imaginary frequencies than it is for real frequencies.

Second, the NRG self-energy also faces issues with $\mathrm{Im}\Sigma(0)$, which is why we cannot use it here to directly benchmark the QMC at low frequencies. It is known that NRG has limited resolution for $|\omega| < T$ \cite{Lee2016}. Therefore, one typically switches from log-Gaussian broadening at $|\omega|>T$ to linear broadening at $|\omega|<T$ to obtain smooth results \cite{Lee2016}. Nevertheless, results such as $\mathrm{Im}\Sigma(\omega)$ for $|\omega| \ll T$ in challenging multiorbital FLs suffer from inaccuracies (say, on the order of 10\%). Again, it is thus advisable to incorporate the FL scaling in the analysis and extract $C$ not just from $\mathrm{Im}\Sigma(0)$, but from an extended frequency range including the $\omega^2$ behavior.

\begin{table}[]
\renewcommand*{\arraystretch}{1.4} 
\begin{ruledtabular}
\centering
\begin{tabular}{l|c|c|c}
& $C$ (eV$^{-1}$) & $\Phi(\epsilon_{\text{F}})$ (eV$/\Omega$cm) & $A$ ($10^{-5}\mu\Omega\text{cm}/\text{K}^2$) \\ 
\hline
SrVO$_3$ & $0.45\pm0.15$ & 2.247 & $3.6\pm1.2$ \\
SrMoO$_3$ & $0.48\pm0.13$ & 3.930 & $2.2\pm0.6$ \\
PbMoO$_3$ & $0.46\pm0.14$ & 3.704 & $2.2\pm0.7$ \\
\hline
SrRuO$_3$ &$\sim10^3$ \footnote{Calculated at $T=0$ K with NRG in the putative paramagnetic phase.} & 2.862 & $\sim0.06\times 10^5$ \\
\end{tabular}
\end{ruledtabular}
\caption{$C$ coefficients extracted by fitting the QMC and NRG real-frequency self-energies to the FL form [Eq.~\eqref{eq:IMSig_FL}] in the range $10\leq\omega/T\leq20$ (see Ref.~\onlinecite{CompanionPaper} and App.~\ref{app:pmo} for more details), value of the transport function at the Fermi level [Eq.~(\ref{eq:Phi})], and $T^2$ coefficient of the resistivity from Fermi-liquid electron-electron scattering [Eq.~(\ref{eq:rho_FL})]. All calculations are in the $Pm\overline{3}m$ paramagentic structure.}
\label{tab:Cdata}
\end{table}

Finally, analyzing the self-energy data on both the imaginary- and real-frequency axes in the way described above yields consistent results for the $C$ coefficient 
(cf.\ Figs.~2--3 in Ref.~\citenum{CompanionPaper} for \svo{} and \smo{} and Fig.~\ref{fig:pmo-C} for \pmo{}).
Together with the value of the transport function at the Fermi level, we can compute the $A$ coefficient via Eq.~(\ref{eq:rho_FL}), see Table~\ref{tab:Cdata}, from which we obtain the resistivity in the FL regime. 
Alternatively, we can use the NRG self-energy or
the ``corrected''  QMC+Pad\'e $\text{Im}\Sigma(\omega)$ [i.e., with $\eta$ chosen to match the FL scaling of Eq.~(\ref{eq:IMSig_FL}), see Fig.~\ref{fig:pade}(a)] in the full Kubo formula, Eq.~(\ref{eq:Kubo}). We will show in Sec.~\ref{sec:resistivity} that all approaches yield consistent $\rho(T)$.

\section{\label{sec:resistivity}Electron-electron contributions to resistivity in cubic oxides}

\begin{figure}
    \centering
    \includegraphics[width=\columnwidth]{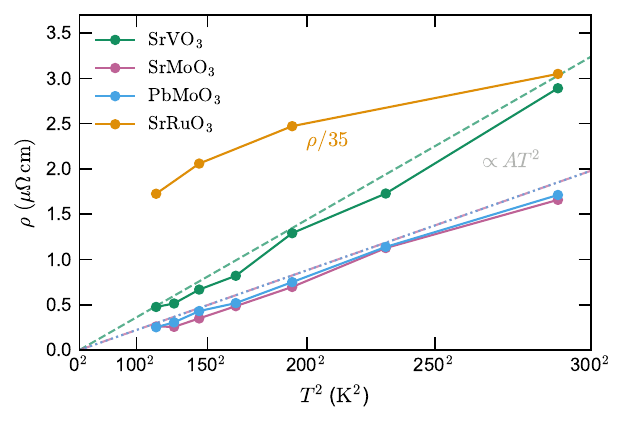}
\caption{Resistivity results from DMFT using the Kubo formula [Eq.~\eqref{eq:Kubo}] for all four materials. Dashed lines indicate $AT^{2}$ obtained from Eq.~\eqref{eq:rho_FL} with $C$ deduced from QMC+Pad\'e and NRG real-frequency data, cf.\ Table~\ref{tab:Cdata}. Note that $\rho$ of SrRuO$_{3}$ is scaled by 1/35.}
\label{fig:rho}
\end{figure}

Using the techniques described in Sec.~\ref{sec:methods}, we present the results for $\rho(T)$ of \svo{}, \smo{}, \pmo{}, and \sro{} in Fig.~\ref{fig:rho}.  As mentioned above, all calculations are for the cubic \cubic{} structure in the paramagnetic phase to facilitate direct comparison. For \smo{}, \pmo{}, and \svo{}, which are in the FL regime at the considered $T$, we compare $\rho$ calculated with the full Kubo formula [Eq.~(\ref{eq:Kubo})] and the simplified FL version [Eq.~(\ref{eq:rho_FL}), extracted using the real-frequency QMC+Pad\'e and NRG data, see Table~\ref{tab:Cdata}]. We observe excellent agreement. Importantly, this agreement is facilitated by the scaling analysis described in Sec.~\ref{sec:trans_FL}. The full Kubo formula, which is sensitive to Im$\Sigma(\omega)$ at low frequency, has a significant dependence on the $\eta$ parameter of the Pad\'e analytic continuation, and thus must be tuned to be consistent with the more reliable higher-frequency region via knowledge of the FL scaling.

A detailed comparison between our results and experiments for \smo{} and \svo{} is provided in our companion paper~\cite{CompanionPaper}. We note that our $\rho(T)$ is remarkably similar between \pmo{} and \smo{}. We thus expect that similar low-$T$ resistivity as in \smo{} or \svo{} will be measured in \pmo{} if high-quality samples (e.g., single crystals or films) can be synthesized and transport can be measured at low enough $T$. 
Also, as discussed in Ref.~\citenum{CompanionPaper}, the magnitude of el-el scattering in \svo{} and \smo{} compared to experiment indicates that el-ph scattering governs $\rho$ at RT; we expect the same to be true for \pmo{}.

\begin{figure}
    \centering
    \includegraphics[width=\columnwidth]{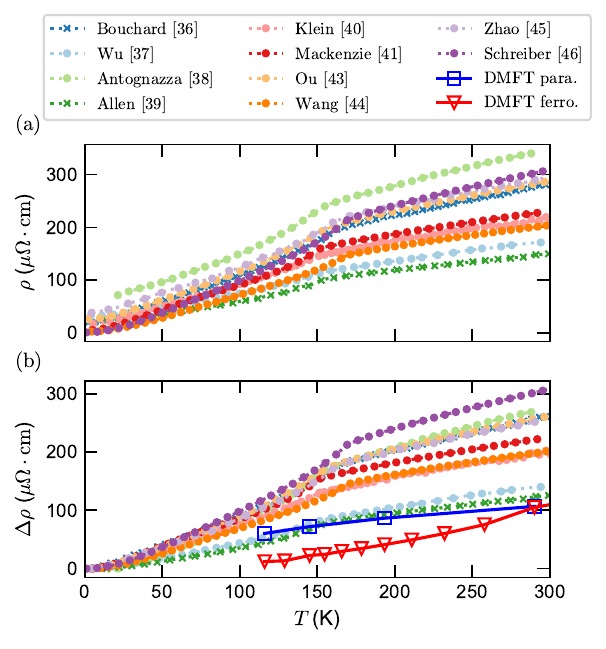}
\caption{Experimental resistivities for \sro{} thin films (circles) and single crystals ($\times$) digitized from Refs.~\citenum{Bouchard1972,Wu1993,Antognazza1993,Allen1996,Klein1996,mackenzie1998,Cao2008,Ou2019,Wang2020,Zhao2021,Schreiber2023} with the residual resistivity (defined by $\rho$ at the lowest extracted $T$ point) subtracted, along with our DFT+DMFT $\rho(T)$ in the paramagnetic (PM) and ferromagnetic (FM) phases.}
\label{fig:sro_exp}
\end{figure}

Finally, we comment on the results for \sro{} in comparison with experiment. In Fig.~\ref{fig:sro_exp}, we plot experimental resistivities for \sro{} thin films and single crystals \cite{Bouchard1972,Wu1993,Antognazza1993,Allen1996,Klein1996,mackenzie1998,Cao2008,Ou2019,Wang2020,Zhao2021,Schreiber2023}, subtracting the residual resistivity (determined by $\rho$ of the lowest extracted $T$ point), compared to our DFT+DMFT results (using QMC). In addition to the paramagnetic results (reproduced from Fig.~\ref{fig:rho}), we include calculations that allow for ferromagnetic order. The latter shows a kink in $\rho$ at 290~K, which is roughly our calculated $T_c$  
(as discussed above, it is expected that single-site DMFT overestimates $T_c$ by roughly a factor of two due to the lack of nonlocal correlations). Once in the ordered phase, $\rho$ is substantially lower than the paramagnetic result, consistent with the trends seen experimentally.

From Fig.~\ref{fig:sro_exp}, it is clear that there is a significant spread in $\rho$ at RT independent of sample quality. It should be noted that Ref.~\citenum{Deng2016} found good quantitative agreement between the DMFT $\rho$ and the results of Klein \textit{et al.}~\cite{Klein1996}. We attribute the fact that we find somewhat lower $\rho$ at RT to a different choice of correlated active space and interaction parameters. In any case, the spread in experimental results makes drawing any quantitative conclusions questionable. However, we can say qualitatively that the DFT+DMFT $\rho$ at RT is on the same order as the experimental measurements, indicating the importance of el-el scattering in this $T$ range, in contrast to the higher-conductivity materials \svo{} and \smo{} \cite{CompanionPaper}. The fact that it is on the lower end of the experimental range does suggest that el-ph scattering may also make a significant contribution for this material.

We can also compare to experiment the calculated low-temperature behavior of \sro{} extracted from the NRG results in Sec.~\ref{sec:FL}.
The $T=0$ K FL behavior of \sro{} follows $-\mathrm{Im}\Sigma = C \omega^2$ with $C \approx 10^3\,$eV$^{-1}$.
With $\Phi(\epsilon_{\mathrm{F}}) \approx 2.862\,$eV/$\Omega\,$cm, 
this corresponds to $A = 24 C / \Phi(\epsilon_{\mathrm{F}}) \approx 0.06\,\mu\Omega\,$cm/K$^2$, which is three orders of magnitude larger than the $A$ coefficient of \svo{} and \smo{}~\cite{CompanionPaper}. Experimental $A$ coefficients obtained from fitting the low-$T$ $\rho$ reported in Refs.~\citenum{mackenzie1998,Cao2008,Wang2020,Schreiber2023} are in the range $A=0.010$--$0.023$ $\mu\Omega\,$cm/K$^2$. Thus, they are similar but somewhat smaller than our value. One difference between experiment and theory is that we calculate in the cubic structure (to facilitate comparisons between oxides); the experimental orthorhombic structure is expected to be slightly more correlated. Hence, this cannot explain the lower $A$ value we obtain compared to experiment. 

Instead, we believe that much of the discrepancy is due to the fact that our NRG calculations were performed in the paramagnetic phase, and \sro{} is ferromagnetic at low $T$. 
As discussed above (see Fig.~\ref{fig:sro_exp}), $\rho$ is significantly larger in the paramagnetic phase, and, indeed, our calculations yield a larger $A$ and a smaller $T_{\text{FL}}$. We can also compare to CaRuO$_3$, which is paramagnetic down to the lowest $T$ measured. The $A$ coefficient of CaRuO$_3$ is somewhat larger ($0.1$--$0.2$ $\mu\Omega\,$cm/K$^2$) \cite{Schneider2014}, which is consistent with the stronger correlations in that material \cite{Dang2015}. We have not attempted computing the $A$ coefficient in the ferromagnetic phase of \sro{}, which would be necessary to quantify the effect of magnetic structure versus, e.g., nonlocal correlations and vertex corrections; the reason is that reaching the required very low $T$ is challenging for QMC and lifting the SU(2) spin symmetry makes multi-orbital NRG computations less efficient and/or accurate. 

\section{\label{sec:conclusions}Summary and Outlook}

Using cubic ABO$_{3}$ perovskites as examples, we demonstrated a DFT+DMFT methodology for accurately extracting the local electron-electron contribution to the dc resistivity in high-conductivity, moderately correlated materials. As for any material with a complex Fermi surface, converged Brillouin-zone integration is crucial for numerical precision. We demonstrate that adaptive integration methods are a powerful systematic way to achieve such convergence of the transport function. The key challenge specific to high-conductivity materials is that the magnitude of the low-$T$ scattering rates are very small. This requires special care when such quantities are extracted from the DMFT self-energies. We show that both low-frequency extrapolations of the Matsubara self-energies \emph{and} Pad\'{e} analytic continuation do not give accurate transport properties. Instead, we advocate leveraging the fact that these materials have moderate correlations, thus relatively high FL coherence scales. The known low-order frequency and temperature scalings can be exploited to extract accurate scattering rates. These scalings also allow QMC calculations at relatively high $T$ to be used to infer transport properties at very low $T$. Comparing results from complementary impurity solvers, e.g., ones that work in real frequencies (like NRG) or imaginary frequencies (like QMC), provides a strategy for validating the consistency of the analysis.

We expect that similar strategies of utilizing FL properties will be useful for calculating other transport properties, e.g., the Seebeck coefficient, Hall conductivity, magnetoresistance, and optical conductivity. Moreover, by ensuring that the numerical aspects of the calculation are under control, this work sets the stage for a careful analysis of the theory of transport in correlated materials. Specifically, 
the role of nonlocal correlations and vertex corrections to the dc conductivity in a framework beyond single-site DMFT remains to be investigated in the context of real materials. Previous works at the model level have shown that vertex corrections are important for low-dimensional systems \cite{Brown2019,Vucicevic2019,Vranic2020,Vucicevic2023} and the low-density limit \cite{Mu2022,Mu2024}. 

Finally, we show for \sro{}
that the methods described here provide results that are qualitatively consistent with experimental resistivity, although such a comparison is challenging due to the spread in reported values and the difficulty of measuring small $\rho$ at low $T$ (this same problem occurs for \svo{} and \smo{}, see Ref.~\citenum{CompanionPaper}). Moreover, the comparison between $\rho$ of \svo{}, \smo{}, and \pmo{} indicates that the superlative conductivity of \smo{} is not a result of, e.g., weaker electron-electron correlations, and must also have its origins in electron-phonon scattering. Thereby, the calculations enabled by our methodology provide significant insight into the transport properties of high-conductivity materials.

\acknowledgements 
We acknowledge useful discussions with Andy Millis, Jernej Mravlje, and Jennifer Coulter. C.E.D.\ and J.L.H.\ acknowledge support from the National Science Foundation under Grant No.~DMR-2237674. F.B.K.\ acknowledges funding from the Ministerium f\"ur Kultur und Wissenschaft des Landes Nordrhein-Westfalen (NRW-R\"uckkehrprogramm). The Flatiron Institute is a division of the Simons Foundation. The data that support the findings of this article are openly available \cite{Article_data}.

\appendix

\section{Crystal data and density-functional theory \label{app:dft}}

DFT calculations are performed using VASP \cite{Kresse:1993bz,Kresse:1996kl,Kresse:1999dk} with the exchange-correlation functionals of Perdew et al.~\cite{Perdew:1996iq}. For these calculations, the following valence states are treated explicitly: Sr ($4s,5s,4p,4d$), Mo ($4s, 5s, 4p, 4d, 4f$), Ru ($4s,5s,4p,4d,4f$), V ($3s,4s,3p,3d,4f$), and O ($2s, 2p, 3d$). The projector-augmented wave method \cite{Blochl:1994dx} is used to treat the core electrons. All DFT calulations are performed in the \cubic structure with $21 \times 21 \times 21$ $k$-point mesh and a cutoff of $550$ eV. The lattice constants read \svo: 3.86 \AA, \smo: 4.00 \AA, \pmo: 4.02 \AA, and \sro: 3.87 \AA~, which were determined by optimizing the unit cell using VASP.

\section{Constrained RPA}
The screened Coulomb interaction is calculated using the implementation of the cRPA in VASP \cite{Merzuk2015}, which uses MLWFs \cite{PhysRevB.80.155134} obtained from Wannier90~\cite{Mostofi_et_al:2014}. For this calculation we used a $7 \times 7 \times 7$ $k$-point grid in the \cubic cubic structure with a cutoff of $500$ eV. The interaction parameters are extracted by averaging the full four index interaction tensor to parametrize the Hubbard-Kanamori Hamiltonian~\cite{vaugier2012}.

\begin{figure}
    \centering
    \includegraphics[width=\linewidth]{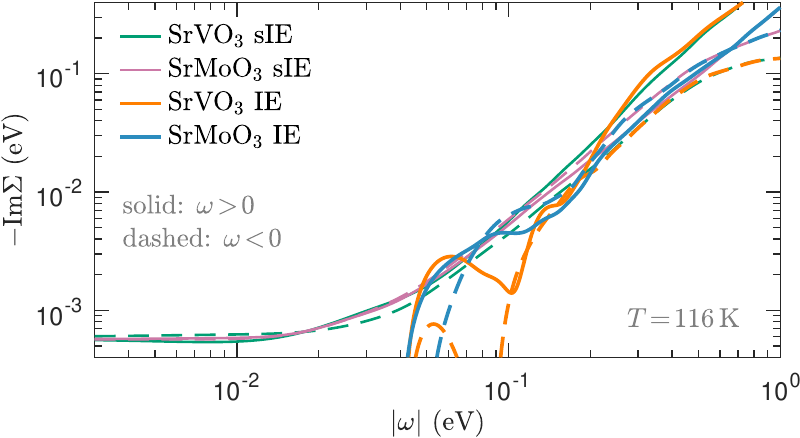}
    \caption{NRG self-energy at $T=116$~K for \svo{} and \smo{}, obtained with the symmetric improved estimator (sIE) from Ref.~\citenum{Kugler2022} compared to the previously used improved estimator (IE) from Ref.~\citenum{Bulla1998}.}\label{fig:Sigma_IE}  
\end{figure}

\section{QMC (TRIQS/cthyb)}
\label{app:QMC}
DMFT calculations were performed using solid\_dmft~\cite{Merkel2022} built on top of the \textsc{TRIQS/DFTtools} software package~\cite{aichhorn_dfttools_2016, parcollet_triqs_2015}. The effective impurity problems are solved using the continuous-time quantum Monte-Carlo hybridization expansion solver as implemented in \textsc{TRIQS/cthyb}~\cite{Seth2016274}. We solve the impurity problems varying temperature from $T=116$ K to $1200$ K. To minimize the Monte-Carlo error in the self-energy, we tune the number of Monte-Carlo measurements such that the auto-correlation $\sim 1$ and directly measure the high-frequency tail of the self-energy~\cite{Labollita2024}. Pad\'e analytic continuation is used to obtain the real-frequency data as described in the main text.

\begin{figure}
    \centering
    \includegraphics[width=\columnwidth]{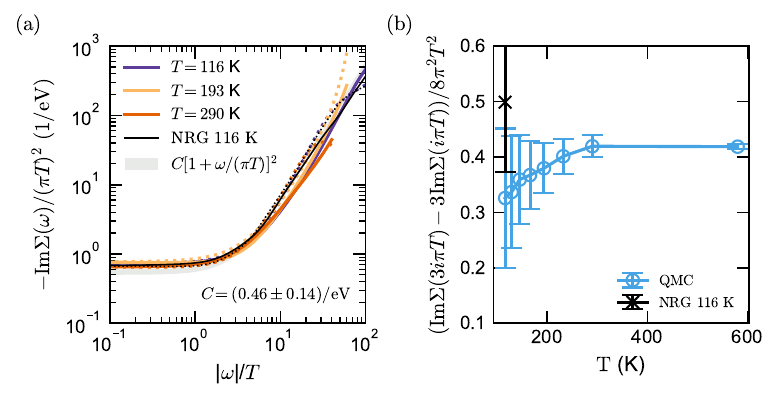}
    \caption{Scattering-rate analysis of the electronic self-energy for PbMoO$_{3}$ in (a) real and (b) Matsubara frequencies from QMC and NRG. (a) $-\mathrm{Im}\Sigma(\omega)/(\pi T)^{2}$ collapses onto the FL form $C[1 + (\omega/\pi T)^{2}]$ which is shown as a gray shaded region. (b) $C$ estimated from Eq.~\eqref{eq:SigMats_FL_formulas} as a function of $T$.}
    \label{fig:pmo-C}
\end{figure}

\section{NRG}
\label{app:NRG}
As another impurity solver,
we use the full density-matrix NRG \cite{Peters2006,Weichselbaum2007} in an implementation \cite{Lee2021} based on the QSpace tensor library \cite{Weichselbaum2012a,*Weichselbaum2012b,*Weichselbaum2020}. 
We express the Hamiltonian in U(1)$\otimes$SU(2)$\otimes$SO(3) symmetric form \cite{Horvat2016},
fully incorporating spin-flip and pair-hopping terms in the interaction.
For a system with $M=3$ orbitals solved with NRG, 
this symmetry is rather low
compared to U(1)$\otimes$SU(2)$\otimes$SU(3) \cite{Stadler2015,Kugler2018,Kugler2024}
or $[\otimes_{m=1}^M$U(1)$]\otimes$SU(2)
\cite{Kugler2020,Kugler2022a,Kugler2024,Grundner2024a},
and thus computationally challenging,
cf.\ Ref.~\citenum{Grundner2024}.
Consequently, we take a rather large discretization parameter $\Lambda = 8$,
keeping up to $N_{\mathrm{keep}}=18.000$ SU(2)$\times$SO(3) multiplets during the iterative diagonalization,
and employ $z$-averaging \cite{Zitko2009} with $n_z = 6$ as well as adaptive broadening \cite{Lee2016,Lee2017}.
Furthermore, the symmetric improved self-energy estimator \cite{Kugler2022} is crucial for reliably extracting the FL scattering-rate parameter $C$ from $\mathrm{Im}\Sigma(\omega)$, see Fig.~\ref{fig:Sigma_IE}.

\section{\label{app:pmo}Analysis of $C$ for PbMoO$_{3}$}
Figure~\ref{fig:pmo-C} summarizes the FL analysis of the electronic self-energy in both Matsubara and real frequencies for PbMoO$_{3}$. On the real-frequency axis (Fig.~\ref{fig:pmo-C}(a)), the self-energies $\Sigma(\omega, T)$ collapse to the form $C[1+(\omega/\pi T)^{2}]$. The coefficient $C$ can then be obtained by fitting the self-energy data to this form. The shaded region denotes the width of the confidence region for the fit. Similarly, on the Matsubara axis, we can use Eq.~\eqref{eq:SigMats_FL_formulas} to obtain an estimate of $C$. Note that, because of the $T^{2}$ in the denominator, we effectively probe beyond the fourth digit of the self-energy, which is difficult to obtain numerically.

\bibliography{bibfile.bib}

\end{document}